\documentclass[manuscript,screen]{acmart}

\usepackage{wrapfig}
\usepackage{float}

\AtBeginDocument{%
  \providecommand\BibTeX{{%
    \normalfont B\kern-0.5em{\scshape i\kern-0.25em b}\kern-0.8em\TeX}}}

\setcopyright{rightsretained}
\copyrightyear{2022}
\acmYear{2022}
\acmDOI{}

\acmConference[KidRec 6]{6th International and Interdisciplinary Perspectives on Children \& Recommender and Information Retrieval Systems (KidRec)
Information Retrieval Systems for Children in the COVID-19 Era; co-located with ACM IDC}{June 27,2 022}{Braga, Portugal}

\acmBooktitle{6th International and Interdisciplinary Perspectives on Children \& Recommender and Information Retrieval Systems (KidRec)
Information Retrieval Systems for Children in the COVID-19 Era; co-located with ACM IDC,
  June 27, 2022, Braga, Portugal}

\begin{document}

\title{Designing Conversational Robots with Children during the Pandemic}



\author{Thomas Beelen}
\email{t.h.j.beelen@utwente.nl}
\affiliation{
 \institution{University of Twente}
 \country{the Netherlands}
 }

\author{Ella Velner}
\email{p.c.velner@utwente.nl}
\orcid{0002-9044-577X}
\affiliation{
 \institution{University of Twente}
 \country{the Netherlands}
 }

\author{Roeland Ordelman}
\email{r.j.f.ordelman@utwente.nl}
\orcid{}
\affiliation{%
  \institution{University of Twente}
  \city{Enschede}
  \country{The Netherlands}
}

\author{Khiet P. Truong}
\email{k.p.truong@utwente.nl}
\affiliation{%
  \institution{University of Twente}
  \city{Enschede}
  \country{The Netherlands}
}

\author{Vanessa Evers}
\email{vanessa.evers@ntu.edu.sg}
\affiliation{
 \institution{NTU Institute of Science and Technology for Humanity}
 \city{Singapore}
 \country{Singapore}
}

\author{Theo Huibers}
\email{t.w.c.huibers@utwente.nl}
\orcid{0002-9837-8639}
\affiliation{%
  \institution{University of Twente - Wizenoze}
  \city{Enschede}
  \country{The Netherlands}
}

\renewcommand{\shortauthors}{Beelen, et al.}

\begin{abstract}
Our research project (CHATTERS) is about designing a conversational robot for children's digital information search. We want to design a robot with a suitable conversation, that fosters a responsible trust relationship between child and robot. In this paper we give: 1) a preliminary view on an empirical study around children's trust in robots that provide information, which was conducted via video call due to the COVID-19 pandemic. 2) We also give a preliminary analysis of a co-design workshop we conducted, where the pandemic may have impacted children's design choices. (3) We close by describing the upcoming research activities we are developing. 


\end{abstract}


\keywords{}

\maketitle

\section{Introduction } 

Children can access digital information in a variety of ways, for example by using a Search Engine (SE) on a computer, or by using a Voice Assistant (VA) 
on a mobile phone, or a dedicated device. As we outlined in \cite{beelen_does_2021}, many of these tools are flawed and do not support children adequately in finding the information they need. 
Our project called CHATTERS\footnote{\url{https://chatters-cri.github.io/}} focuses on a physically embodied agent (robot) as a search tool for children of 10-12 years old \cite{beelen_does_2021}. The project was defined in 2019 before the COVID-19 pandemic.  The physical form of this agent provides an engaging interaction, as well as drawing attention in the context of a museum. Similarly to voice agents, our robot uses speech to communicate with a child. The project consists of two research topics: conversational search for children, and how the robot can make sure this happens responsibly. Where commercial Voice Agents use a limited interaction model consisting of a query and a direct response \cite{lovato_hey_2019}, we suggest a spoken conversational interaction. This interaction instead follows the Spoken Conversational Search paradigm, meaning there is a mixed-initiative, back and forth around information needs \cite{zamani_conversational_2022}. To better support children, we study how we can use search via responsible conversation.

While a robot may help children in their search journey by using speech and conversation, we also need to consider the risks of children using such an embodied agent. Robots are prone to errors. Furthermore, autonomously moderating online information is infeasible. This may lead to a robot presenting false or misleading information. Knowing that children build social bonds with robots and are prone to trust them, this could create precarious situations \cite{Belpaeme2013Multimodal2, DiDio2020ShallRelationships}. To monitor how children assess the robot and the information it provides, we propose using the child's speech during the interaction to measure their trust in the robot. If this turns out to be too high (which is often referred to as overtrust \cite{Lee2004TrustReliance, deVisser2019TowardsTeams}), the robot should intervene by altering its behavior to attempt to lower the trust. This can be seen as a \textit{sense-think-act loop}, where the robot senses the trust level of the child in the robot, assesses whether there is overtrust, and acts accordingly with an intervention. 

We address the two research tracks by multiple experiments and (participatory) design activities. The next sections describe an experiment on children's perception of robots and online information, as well as a co-design activity that has been conducted. We focus on the impact of the global COVID-19 pandemic on search technology for children, as well as the impact on research methods. 

\section{Empirical study: Children's view on robots and digital information} \label{exp1}
\begin{figure}[t]
    \centering
    \includegraphics[width=\textwidth]{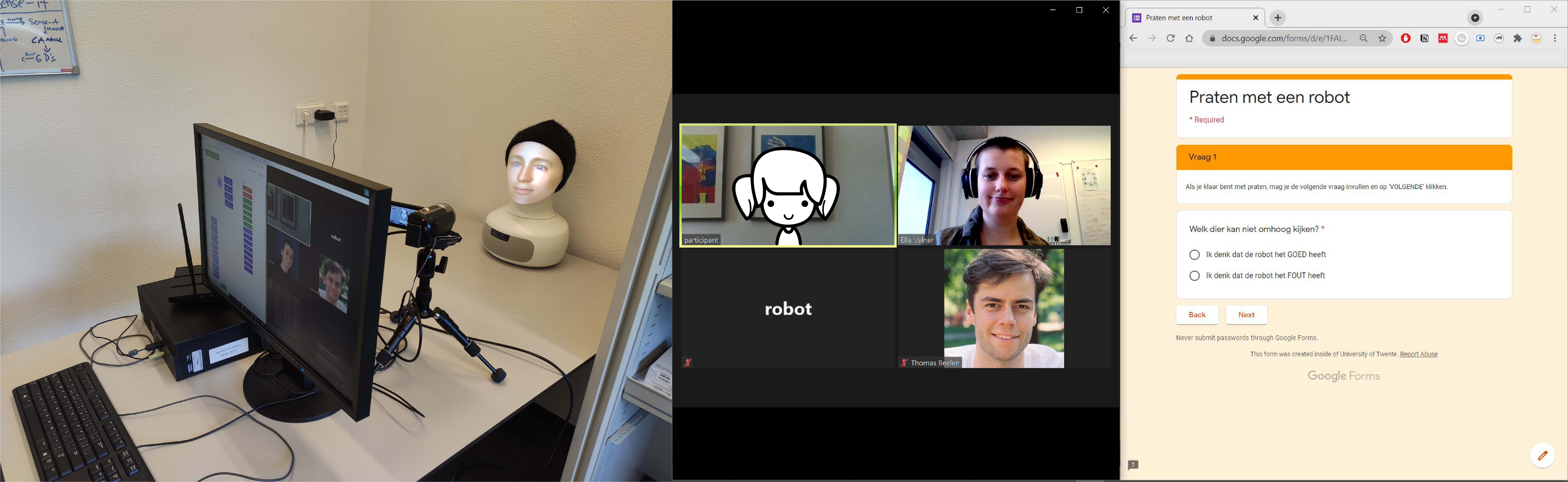}
    \caption{Setup of experiment 1, where the robot was in a separate room with a camera, and appeared whenever the child was interacting with it. The child had a laptop with on the left side of the screen the video call, and on the right side the questionnaire.}
    \label{fig:setup}
\end{figure}

A first experiment was designed to 1) gain knowledge on how children's trust in a robot influences their attitude and behavior towards the robot and the information it provides, and 2) to collect speech data of children interacting with a robot in high and low trust conditions. In June 2021, we recorded 30 children (10-12 years old) playing a quiz with a trustworthy and untrustworthy robot and measured their attitudes toward the robots with questionnaires (trust, likability, knowledgeability, intelligence). Because of COVID-19, the experiment was done via video call (see Fig. \ref{fig:setup} for setup) to minimize face-to-face interactions. After children interacted with both robots, we conducted a semi-structured interview. The goal of the interview was to gain insight on how children perceive the robots, how they expect robots to deal with credibility of online information, and their willingness to use a robot. Additionally, we asked them about their current use of voice assistants since this is a widespread technology for searching with voice. Although this study revealed many interesting results, we want to highlight the results of the interview since these are particularly interesting for the community of this workshop. A more detailed description of this study and other results will be reported on in a later publication.

\subsection{Results}
The participants (N=30) were asked about their view on robots, the information they can provide, and their current voice agent usage. Halfway throughout the study, we realized it is also valuable to ask children about their view on interacting with a robot over video-call compared to face-to-face (N=15). All child answers are translated from Dutch.

\subsubsection{Robots and the digital information}
Children seem to be aware of the existence of ``bad information'' and ``bad websites''. When asked what the robot should do with this, they suggest using specific websites that are known to be knowledgeable on the topic, or to compare multiple websites. One child also suggested that robots should use websites based on search engine ranking. However, not all children think robots are (currently) able to recognize these ``bad websites''. One child stated that robots are still in development and might be able to do this in the future. Several children mention the programmer of the robot, and that they can program the robot to be able to make this distinction, revealing a fairly profound understanding of how robots are made.

\subsubsection{Current voice agent usage}
Since currently spoken search is done with voice agents, such as Google Home or Siri, we would like to know if children are actually using these agents for this purpose, as previous research suggests \cite{lovato_hey_2019, garg_he_2020}. Two children stated that they don't trust voice agents and therefore don't use them, because they might be able to listen in or record random conversations. Some children do not use voice agents, stating that typing is easier. However, others confirmed previous work that voice agents might be preferred for children as opposed to traditional ways of search \cite{druin_how_2009, jochmann-mannak_children_2010}. They stated it ``is easier than typing'', and they use it ``when I don't know how to write it''. Most children that use voice agents use it for fun (e.g., music, jokes, funny interactions). Some also stated that they use it for information or to call or write messages to people, especially ``when you have your hands full''. 

\subsubsection{Video call compared to face-to-face}
Since talking to a physical robot in the same room might give a different experience than talking with a robot in front of a camera \cite{li_benefit_2015}, 15 participants were asked how they think the video-mediated interaction compares to a face-to-face interaction with the robot. Note that this was a hypothetical question for most children, since they had never interacted with a robot face-to-face. Children's thoughts on these similarities and differences could especially be valuable with possible future lock-downs in mind. Furthermore, performing HRI studies online via video call could give more opportunities for much needed intercultural studies.

Many children think it would be different to talk to a robot face-to-face, as expected. 6 children said this would be more fun, but 5 children also stated that it might make them more nervous. Although nervousness does not necessarily have to be a bad thing, it might be interesting to study this further and investigate where the nervousness comes from. One child explained this as ``just like talking to the king over zoom is different than in real life''.

\section{Co-Design workshop: Children designing a conversational robot}
The robot we are designing needs to fit into children's lives. This introduces several design challenges such as the robot's physical design, social role, personality traits, conversational style, and approach to online information credibility. We invited children to a workshop (Februari 2022) as \textit{design-partners} to work with us on these challenges \cite{druin_role_2002}. 

During the workshop, children worked in groups on a worksheet with questions, robot design, and a storyboard. The worksheet is titled \textit{"Design your own homework robot"}, and we describe the robot may help the user by finding information. It has a few questions that prompt the children to discuss design aspects (e.g. characteristics and skills) together, a drawing of the robot, and finally a storyboard bringing it all together and demonstrating some of the robots capabilities. We worked together with a Dutch school to run these workshops, that also included an interactive presentation and a robot programming activity. To fit our activities into a school day, we needed to work with an entire class at once. Both two researchers and a teacher walked around the groups to assist where possible. Parents were asked for consent to include children's worksheets in our analysis.

\begin{wrapfigure}{!h}{0.30\textwidth}
    \centering
    \vspace{-0.3cm}
    \includegraphics[width=0.30\textwidth]{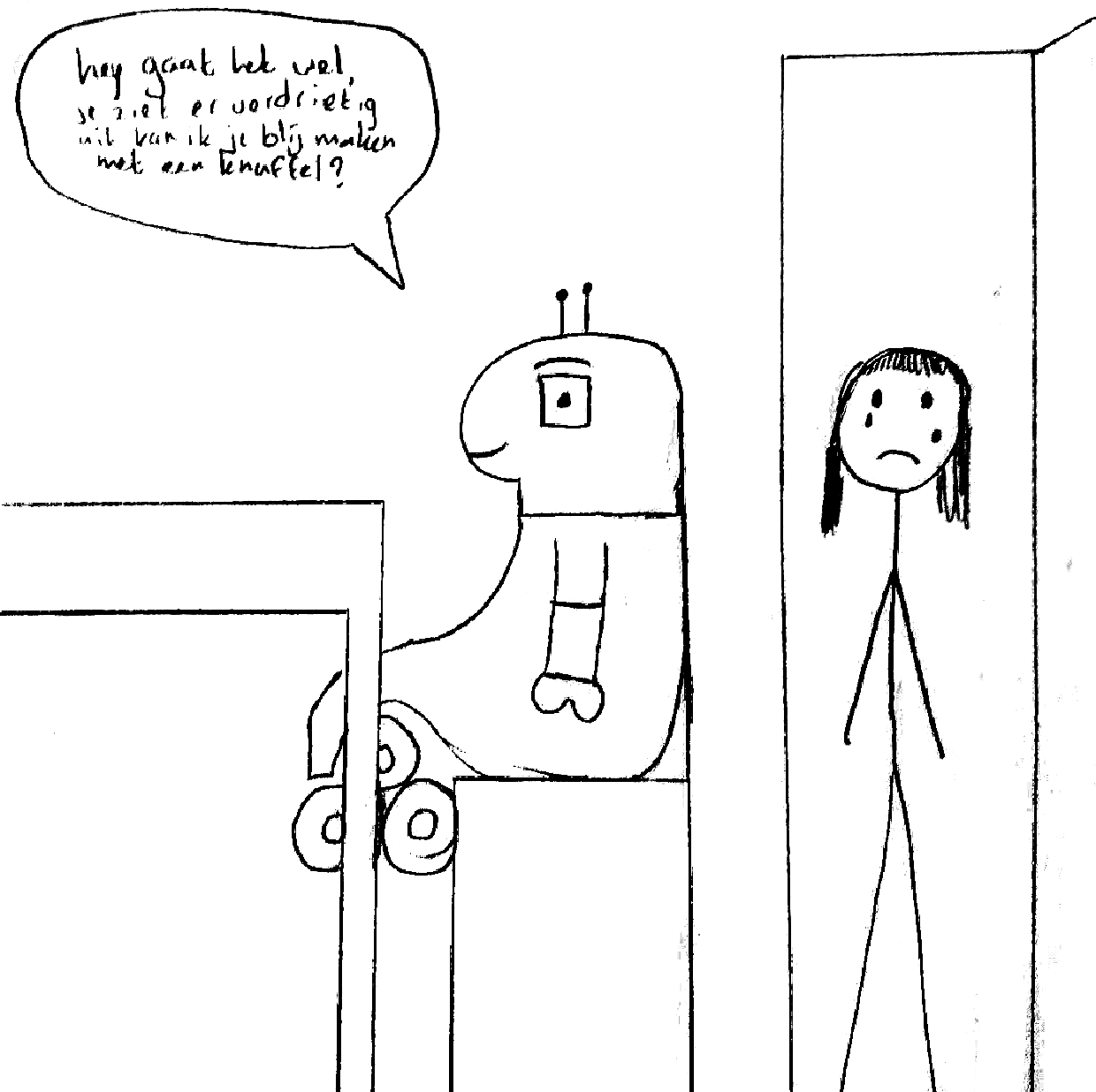}
    \caption{Excerpt from a storyboard. Robot (translated): \textit{"Hey are you okay? You look sad. Can I cheer you up with a hug?"}
    \vspace{-0.5cm}
    }
    \label{fig:hug}
\end{wrapfigure}

\subsection{Results}
In total, the results of 69 children are included in our analysis, constituting to 18 design groups (3-4 children per group). The worksheets were analysed by two researchers independently and compared afterwards. Differences were discussed and resolved. The full analysis of the workshop is to be reported in a later publication. In this document we provide preliminary results with a focus on COVID-19 related outcomes. 

\paragraph{Mobility}
Most groups designed a robot for at home (13 groups), this may have been influenced by framing the design challenge as a \textit{homework robot}. Many groups designed robots that are mobile or portable and can thus be used at multiple locations. Many robots have legs or wheels, and in some cases it is explicitly mentioned that the robot is able to ascend stairs for example. This may be reflective of changes due to COVID-19, such as increased working from home where school materials may be taken home. Remarkably, several groups mention or depict the robot charging it's batteries. Some even show a robot that can plug itself in to charge. This attention to charging and battery power seems to further emphasize the need for a mobile or portable system that can be used in multiple locations. This focus does not seem present in earlier comparable design sessions \cite{landoni_youve_2020}.

\paragraph{Companionship}

More than half of the groups personified their robot by giving it a name, in line with preferences discovered in previous work \cite{yarosh_children_2018}. Furthermore, many groups emphasize the social character of the robot and focus on companionship. The five most used words to describe the robots' character are (translated from Dutch, in descending order): \textit{smart, funny, helpful, kind,} and \textit{social}. One group's storyboard begins with the robot noticing the child is in a bad mood and offering the child a hug as shown in figure \ref{fig:hug}. Another group shows the robot referring to the user as \textit{"bestie"}. Previous work on co-designing search agents with children also shows that children prefer a social and friendly robot \cite{landoni_youve_2020}, though it may be possible that the COVID-19 pandemic increased the need of companionship.

\subsection{Limitations} 
The work we describe comes with several limitations. 1) The children that participated (11-13) were older than our target audience(10-12), where \citet{druin_cooperative_1999} found children of ages 7-10 to be the best design partners. 2) Researchers and teachers had limited time to spend with each design group, while past studies suggest inter-generational work best \cite{druin_cooperative_1999}. However, the fact that children are a bit older in our sample may lead to a greater independence within teams. 
Our design partners being a bit older (limitation 1), as well as limited researcher interaction (limitation 2) may be the cause of children's disinterest in some cases. Some worksheets contained memes or jokes that may be indicative of this.

\section{Future studies}
We will continue our investigations in this project in the two previously described directions: conversational design and how we can do this responsibly, keeping in mind children's experiences and challenges in current times. We are currently developing two studies on the design of the conversation between children and a search agent, as well as a study on the how to further dampen the trust of children in robots. These will be outlined below.

\subsection{Conversational design} 

\subsubsection{Children's search conversations}
In this planned study we want to learn about children's behavior and expectations during conversational search. Inspired by \citet{trippas_towards_2020}'s study with adults, children will search for information in dyads. One of the children takes the role of \textit{intermediary}, the other of \textit{seeker}. The seeker is given a search task, and the intermediary has access to a computer with a search engine. The participants communicate via speech and have to work together to solve the search tasks. The participants are seated in a way that they can see each other, but only the intermediate can see the computer screen. By analysing the resulting conversations, learn more about the child-specific interaction patterns and expectations of conversational search for children. The interactions will be annotated using the SCoSAS annotation scheme \cite{trippas_towards_2020}. Then conversational aspects such as type of utterance (both parties), information requests (seeker), query formulation (intermediary), can be analysed and compared to findings with adults. 

\subsubsection{Conversational compared to query-response search}
The second study we are developing has the goal to see whether an agent using a simple conversational approach is more successful in helping children during the search process. We want to find out if the simplified approach leads to more semantically rich descriptions of the child's information need, and how children experience the interaction. In the simplified approach, the robot aims to elicit semantically rich descriptions of the child's information need by asking questions that are generated with a rule-based algorithm. This algorithm will identify search terms in the child's speech and ask to elaborate. 
The semantic richness in a conversational setting will be compared to an interaction with an agent using a more traditional query-response interaction, commonly found in commercial voice agents. To measure children's experience we will use a questionnaires on engagement and trust, as well as a semi-structured interview. 

\subsection{Children's trust in the search agent}
We are currently analyzing the speech that we collected in the first study described in section \ref{exp1} to search for evidence that trust is reflected in child's speech. Although we tried to manipulate the robot to be either very untrustworthy or very trustworthy, we noticed high trust in both of them. Although even small behavioral changes in a robot is reported to potentially influence a child's perception of the robot~\cite{Peters2017RobotsCompetence}, our manipulation might not have been strong enough to achieve the impact we were aiming for. Alongside the behavior and appearance of the robot (which was the focus of the first manipulation), the robot's context is also a prominent factor on the trust people have in robots \cite{Cameron2015FramingInteraction}.
To look further into manipulating trust levels, we are thinking about purposely misaligning the context of the robot and the interaction. This could be implemented in the envisioned sense-think-act loop to dampen cases of overtrust.

\section{Discussion}
In this document we described our ongoing project on robots for children's IR. We described initial results of two recent studies where our analysis focused mainly on COVID-19. The global pandemic impacted our way of conducting research. Children we asked after the video-mediated robot interaction, expect that their experience changed compared to face-to-face interaction. In our design sessions with children, we noticed a strong focus on mobile robots, as well as companionship. These factors may be influenced by the pandemic where children spent extended periods following school from home. We further outlined the planned research in our project, which was updated with regards to earlier workshop contributions. We thereby continue working towards responsible conversational robots for children's information search.

\section{Acknowledgements}
This research is supported by the Dutch SIDN fund
(https://www.sidn.nl/) and TKI CLICKNL funding of the Dutch Ministry of
Economic Affairs (https://www.clicknl.nl/).

\bibliographystyle{ACM-Reference-Format}
\bibliography{kidrecfile}

\end{document}